\documentclass[%
 reprint,
superscriptaddress,
 amsmath,amssymb,
 aps,
]{revtex4-2}

\usepackage{graphicx}
\usepackage{dcolumn}
\usepackage{bm}

\usepackage{color,amsmath}
\usepackage{physics}
\usepackage{subfiles}

\usepackage[normalem]{ulem}
\usepackage[usenames,dvipsnames]{xcolor}
\usepackage{hyperref}
\hypersetup{
    colorlinks,
    linkcolor={violet},
    citecolor={ForestGreen},
    urlcolor={teal}
}

\newcommand{\beq}{\begin{eqnarray}}
\newcommand{\eeq}{\end{eqnarray}}
\newcommand{\bea}{\begin{eqnarray}}
\newcommand{\eea}{\end{eqnarray}}
\newcommand{\be}{\begin{equation}}
\newcommand{\ee}{\end{equation}}
\newcommand{\bq}{\begin{equation}}
\newcommand{\eq}{\end{equation}}

\newcommand{\half}{\frac{1}{2}}
\newcommand{\nn}{\nonumber}

\def\half{\frac12}

\def\m{\mu}
\def\n{\nu}

\def\6{\partial}

\def\6{\partial}

\begin{document}


\title{An inflationary cosmology from anti-de Sitter wormholes}

\author{Panos Betzios}
\email{pbetzios@phas.ubc.ca}
 \affiliation{\href{https://phas.ubc.ca/}{Department of Physics and Astronomy}, University of British Columbia, \\
6224 Agricultural Road, Vancouver, B.C. V6T 1Z1, Canada}

\author{Olga Papadoulaki}
\email{olga.papadoulaki@polytechnique.edu}
\affiliation{\href{https://www.cpht.polytechnique.fr/?q=en}{CPHT, CNRS, École polytechnique, Institut Polytechnique de Paris}, 91120 Palaiseau, France}

\date{\today}

\begin{abstract}
We propose a new type of wavefunction for the universe computed from the Euclidean path integral, with asymptotically $AdS$ boundary conditions. In the semiclassical limit, it describes a Euclidean (half)-wormhole geometry, exhibiting a local maximum of the scale factor at the surface of reflection symmetry, giving rise to an expanding universe upon analytic continuation to Lorentzian signature. We find that these Euclidean wormholes set natural initial conditions for inflation and that the semi-classical Wheeler-DeWitt wavefunction can favor a long lasting inflationary epoch, resolving a well known issue of the no-boundary proposal. Due to the asymptotic $AdS$ conditions in the Euclidean past they raise the possibility of describing the physics of inflating cosmologies and their perturbations within the context of holography.
\end{abstract}

\maketitle


\section{Introduction}

Research in cosmology has become extraordinarily lively in the last fifty years. Observations by instruments both ground-based and on the sky, and eventually by the Wilkinson Microwave Anisotropy Probe~\cite{Bennett_2013}, have shown a remarkable agreement with the predictions of the inflationary theory~\cite{Guth:1980zm,Linde:1981mu,Linde:1984ir}. One of the main advantages of the inflationary paradigm is that whilst solving the traditional horizon, flatness and monopole cosmological problems, it also explains quite naturally the existence of the primordial cosmological perturbations generated as the result of the enhancement of the inflaton vacuum quantum fluctuations due to the accelerated cosmological expansion~\cite{Mukhanov:1981xt,Starobinsky:1982ee,Guth:1982ec,Bardeen:1983qw}.

However within the inflationary scenario, the question of initial conditions that drive inflation itself is not determined. As one traces the cosmic evolution back in time, the curvatures and matter densities seem to approach the Planck scale, where one expects quantum gravitational effects to come into play. Such manifestly singular loci of spacetime ---spacelike singularities--- seem to suggest that one might need a UV complete theory of quantum gravity to tackle the question of initial conditions. Yet again, one may hope that there exist some (perhaps effective) description of the physics of the very early universe, that evades this UV issue, bringing back such questions within our scientific grasp. Related to this, the quantum state of a spatially closed universe (a three hyperfurface $\Sigma$), is expected to be described by a wave function $\Psi_\Sigma(g_{i j}(\vec{x}), \Phi(\vec{x}))$ which is a functional depending on the geometry and on the values of the matter fields on $\Sigma$. This wave function should obey the Wheeler-DeWitt (WDW) constraint equation~\cite{DeWitt:1967yk} ---the quantum analogue of the Hamiltonian constraint of general relativity. 
On the other hand, the WDW equation admits many solutions, so that further input is needed on how to select a subset of them, using appropriate boundary conditions. The no-boundary proposal introduced by Hartle and Hawking~\cite{Hartle:1983ai,Hartle:2007gi},
and the tunneling proposal by Vilenkin~\cite{Vilenkin:1982de,Vilenkin:1983xq}, posit that the universe could have a quantum beginning corresponding, at least semiclassically, to a compact Euclidean geometry. Observations suggest that the early universe was simpler than it is now ---more homogeneous and isotropic--- a characteristic of ground states in physical systems. A natural idea is that the universe should begin in a cosmological analogue of a ground state. If we assume positivity of the inflaton potential for all the field range, the no-boundary proposal seems to be our best candidate consistent with this idea, whilst at the same time predicting a correct (nearly Gaussian) spectrum of primordial perturbations~\cite{doi:10.1142/1190,Lehners:2023yrj,Maldacena:2024uhs}. At the same time, it also gives rise to a great puzzle: it predicts an empty universe with the least possible number of inflationary e-folds~\cite{doi:10.1142/1190,Lehners:2023yrj,Maldacena:2024uhs}.

\begin{figure}[b]
\includegraphics[width=80mm]{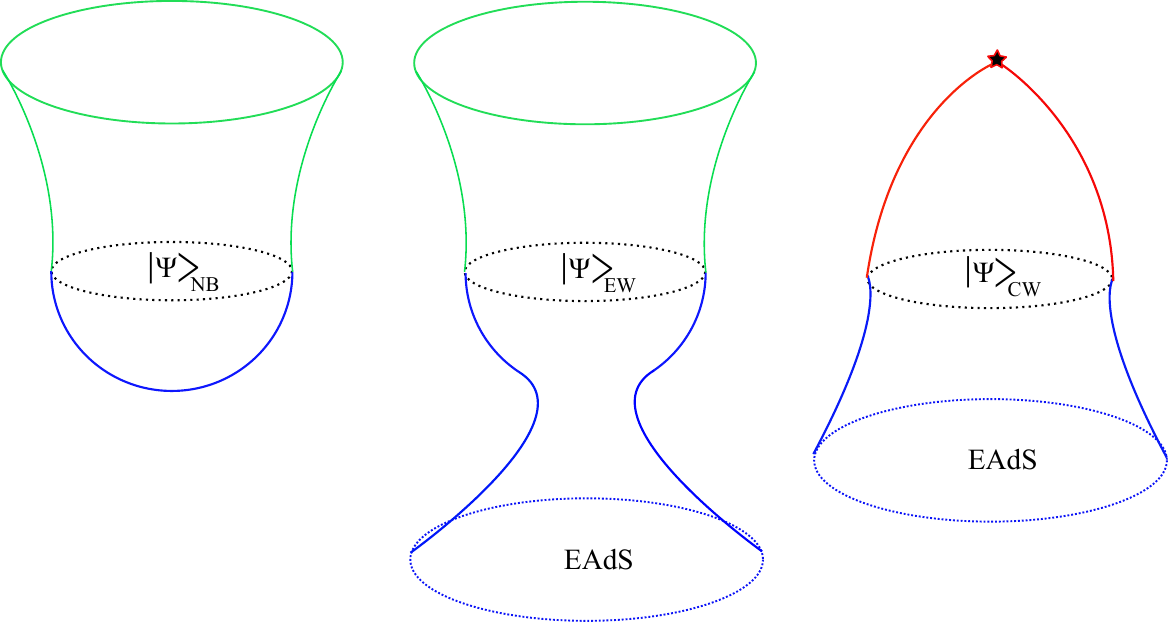}
\caption{\label{fig:wavefunctions} Three types of wavefunctions computed from the Euclidean path integral and their subsequent Lorentzian evolution. The first is an example of the no boundary (NB) proposal ---a half-$S^4$ evolves into a Lorentzian $dS_4$ spacetime. The last one corresponds to a contracting Euclidean $AdS$ half-wormhole, giving rise to a crunching cosmology. In the middle we depict the type of ``wineglass" half-wormholes we are interested in this work ---they combine features of the other two cases by approaching the $EAdS$ vacuum in the far past and expanding near the $\tau=0$ slice, giving rise to an inflating Lorentzian cosmology upon analytic continuation.}
\end{figure}

After the advent of the AdS/CFT correspondence, we now know that theories of quantum gravity are better defined and behaved in the presence of a negative cosmological constant~\cite{Maldacena:1997re,Witten:1998qj,Gubser:1998bc}. The negatively curved anti-de Sitter (AdS) spacetime corresponds precisely to the bulk ground state that is dual to the ground state of a dual CFT. In this setting, it is possible to recover a cosmological FRW spacetime upon analytic continuation of \emph{Euclidean AdS wormhole geometries}~\cite{Maldacena:2004rf,Betzios:2017krj,Betzios:2019rds,Betzios:2021fnm,Antonini:2022blk,Antonini:2022ptt}, but the resulting cosmologies inevitably crunch in the future and do not seem to allow for a sufficient period of inflation~\footnote{Although it has been argued that the aforementioned cosmological puzzles might be resolved by other means~\cite{Antonini:2022ptt}.}.

\begin{figure}[b]
\includegraphics[width=90mm]{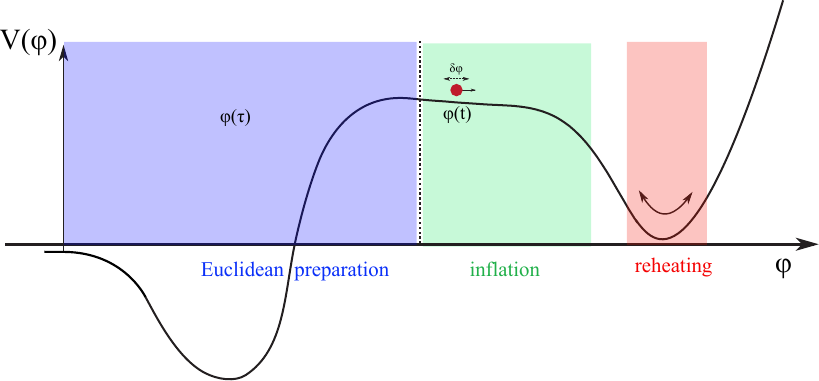}
\caption{\label{fig:PotentialHistory} A typical form for the scalar potential. The vertical dashed line indicates the initial state prepared under Euclidean evolution with asymptotically $EAdS$ boundary conditions (see Figs.~\ref{fig:wavefunctions} and~\ref{fig:winewormhole} for details). This state then evolves and inflates in Lorentzian time $t$ (we depict a typical form of a slow roll potential, with a subsequent reheating phase).}
\end{figure}

In this Letter, we propose that \emph{the initial state of inflation is set forth by a certain kind of Euclidean AdS ``wineglass"  wormholes} depicted in Fig.~\ref{fig:wavefunctions}
and that the issues afflicting the Hartle-Hawking and Vilenkin proposals can be naturally solved in this setting. Our proposal rests upon the following assumption: The effective inflaton potential should admit at least a stable $AdS$ minimum, as well as a metastable $dS$ minimum. From an $AdS/CFT$ point of view, it is natural to consider a model where there is an additional local unstable $AdS$ maximum, which in the language of holography leads to a renormalization group flow driven by a relevant operator, see Fig.~\ref{fig:PotentialHistory}. We should emphasize that it is \emph{a proposal for the pre-inflationary epoch}, that sets natural initial conditions and a reasonable probability weight for a long-lasting inflationary period. The precise subsequent Lorentzian evolution is determined from detailed properties of the inflaton potential, and our construction can accommodate various possibilities, consistent with current experimental data~\cite{Planck:2018jri}.

\section{A Model}

Our proposal can be incorporated in a large class of models as long as they satisfy the general properties put forward in the introduction. For concreteness we focus in a specific simple model of Einstein gravity, minimally coupled to a scalar (inflaton) field, with a potential $V(\phi)$. We shall also consider the presence of an axion field. Axionic fields do appear in string theory and supergravity, as well as in the Peccei-Quinn proposal to deal with the strong-CP problem and are dark matter candidates. Analogous models have been discussed in the literature, see \cite{Gutperle:2002km,Hertog:2017owm,Andriolo:2022rxc} for $EAdS$ axion wormholes and~\cite{Rubakov:1988wx,Jonas:2023ipa} for more similar ``wineglass" shaped wormholes, albeit in asymptotically flat space. In view of phenomenological applications, we focus in the case of four spacetime dimensions, but our model is easily generalisable to an arbitrary number of dimensions. The Euclidean Einstein-scalar action is ($\kappa \equiv M_{Pl}^{-2} = 8 \pi G_N$)
\be\label{Einsteinscalaraction}
S_E = \int d^4 x \sqrt{g_E} \left(-\frac{1}{2 \kappa} R + \half \nabla^\mu \phi \nabla_\mu \phi + V(\phi)   \right) \, ,
\ee 
to which we add an additional axionic contribution
($f_\alpha$ is the axion coupling constant and $H_{\m \n \rho}$ its three-form field strength)
\be\label{axionmatterradiation}
 S_E^{\text{axion}} = \int d^4 x \sqrt{g_E} \frac{1}{12 f_\alpha^2} H_{\m \n \rho} H^{\m \n \rho} \, .
\ee
Assuming the simplest spherically symmetric and homogeneous ansatze ($q$ is a constant axion charge)
\be\label{minisuperansatze}
ds^2 = d \tau^2 + a^2(\tau) d \Omega_3^2 \, , \quad \phi(\tau) \, , \quad  H_{i j k} = q \epsilon_{i j k} \, , 
\ee
we find the set of equations of motion (the axionic one is trivially satisfied), where the prime denotes a $\tau$ derivative 
\bea\label{EEOMs}
\frac{2 a''}{a} + \frac{a'^2}{a^2} - \frac{1}{a^2} + {\kappa} \left(V(\phi) + \frac{\phi'^2}{2} \right) - \frac{\kappa Q^2}{ a^6}  = 0   \, , \nn \\
\frac{a'^2}{a^2} - \frac{1}{a^2} + \frac{\kappa}{3} \left(V(\phi) - \frac{\phi'^2}{2} \right) + \frac{\kappa Q^2}{3 a^6}  = 0 \, , \nn \\
\phi'' + 3 \frac{a' \phi'}{a} - \frac{d V}{d \phi} = 0 \, , \qquad \qquad
\eea
where $Q^2 \equiv  q^2/ 2 f_\alpha^2 $. As is well known the Euclidean ODE for $\phi(\tau)$, describes a particle moving in an effective inverted potential $U_E(\phi) = -V(\phi)$, with a friction term $3 a' \phi'/a$ (this can be anti-friction if $a'/a< 0$), see Fig.~\ref{fig:winewormhole}.

\subsection{Euclidean $AdS$ ``wineglass" wormholes}

\begin{figure*}
\includegraphics[width=150mm]{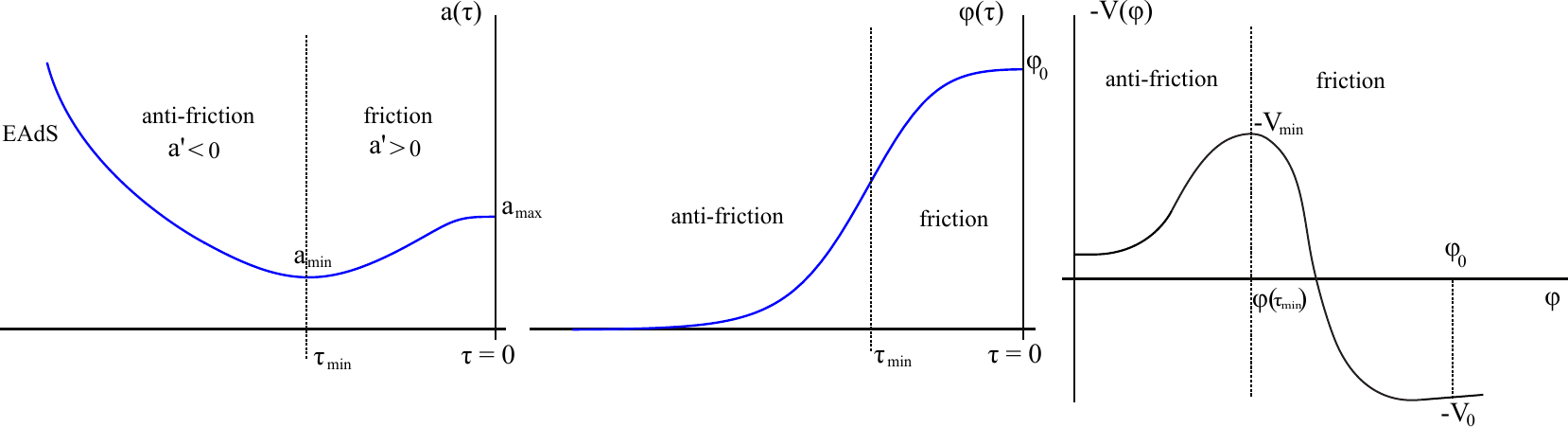}
\caption{\label{fig:winewormhole} The Euclidean evolution of the scale factor and the scalar field for a $EAdS$ ``wineglass" half-wormhole. The two boundary wormhole is found by extending the solution to positive $\tau$ (upon time reflection one has to interchange friction with anti-friction). In the right figure we depict the mechanical analogue of a particle moving in the inverted potential $-V(\phi)$. The dashed line and $\phi(\tau)$ are drawn for the special case where the friction regime starts at the maximum of the inverted potential. 
}
\end{figure*}

The Euclidean ``wineglass" wormhole solutions that we are interested in, have the property that they asymptote to a $EAdS$ space ---$a(\tau) \sim \exp(H_{AdS} |\tau|)$ at $\tau \rightarrow \pm \infty$. Moreover we demand that they satisfy the following conditions at $\tau = 0$
\be\label{initialconditions}
a''(0) < 0 \, , \quad a'(0) = 0 \, , \quad a(0) = a_{\text{max}} \, , \quad \phi'(0) = 0 \, , 
\ee
so that $a_{\text{max}} $ is a local maximum of the scale factor. From the second equation in \eqref{EEOMs}, we also find that $x \equiv a_{\text{max}}^2$ obeys a cubic equation~\footnote{See~\cite{Jonas:2023ipa} and the supplementary material provided for more details.} with its value bounded by
\be\label{boundamax}
\frac{2}{\kappa V(\phi_0)} < a_{\text{max}}^2 \leq \frac{3}{\kappa V(\phi_0)} \, .
\ee
The equality is saturated when the throat acquires the largest possible (Hubble radius) size for vanishing axion charge ($Q = 0$). In addition there exists another point in Euclidean time for which the scale factor acquires a local minimum $a''(\tau_{min}) > 0 , \, a'(\tau_{min}) = 0$. This local minimum, signals a change in the sign of the friction term in
the scalar equation of \eqref{EEOMs}, see also Fig.~\ref{fig:winewormhole}. In our models for concreteness we consider a potential $V(\phi)$ that has a local maximum at $\phi=0$, where the geometry becomes that of $EAdS$ (i.e. $V(\phi) \sim -1 + m^2 \phi^2/2$, in AdS units). In the context of holography (AdS/CFT), this means that there is a renormalization group flow driven by a relevant operator with conformal dimension $\Delta = 3/2+\sqrt{9/4+m^2} < 3$ (there is also a Unitarity or Breitenlohner-Freedman bound constraint so that $-9/4 \leq m^2$). The potential drawn in Fig.~\ref{fig:winewormhole} can support other types of wormhole solutions such as the ones studied in~\cite{Betzios:2019rds,Antonini:2022blk,Antonini:2022ptt,Hertog:2017owm,Andriolo:2022rxc}, but the resulting cosmologies eventually crunch instead of inflate, hence they are not so interesting for phenomenological purposes. One can also envisage interesting wormhole solutions driven by irrelevant operators, with a minimum of $V(\phi)$ at $\phi = 0$, but then the potential/initial conditions should be fine-tuned so that the Euclidean motion in $-V$ does not over/under-shoot what is now a local maximum.

\subsection{Subsequent Lorentzian evolution}

The analytic continuation to Lorentzian signature, is performed by rotating $t = - i \tau$. We denote the real time derivatives by a dot i.e. $\dot{a}$. We first observe that the conditions \eqref{initialconditions} set forth from the Euclidean evolution are consistent with an initially expanding universe ($\ddot{a}(0)>0$) of Hubble radius size ($a^2(0) \simeq 3/\kappa V(\phi(0))$), with vanishing initial kinetic energy for the scalar field ($\dot\phi (0)=0$) ---these serve as perfect initial conditions for inflation. The field then evolves in a slow roll potential such as the one drawn in Fig.~\ref{fig:PotentialHistory}, where we draw a typical ``Hilltop" slow roll model~\cite{Boubekeur:2005zm} supporting a subsequent reheating phase. The validity of the slow roll approximation rests upon assuming small slow roll parameters
\be\label{slowrollpar}
\epsilon_V \equiv \frac{M_P^2}{16 \pi} \left(\frac{V_\phi}{V} \right)^2 \ll 1 \, , \quad  \eta_V \equiv \frac{M_P^2}{8 \pi} \frac{V_{\phi \phi}}{V} \ll 1 \, ,
\ee
for the (light green) region in the potential $V(\phi)$ on the right of the vertical dashed line in
Fig.~\ref{fig:PotentialHistory}. The number of inflationary e-folds $N_*$ is found from integrating $d N \simeq  d\phi/M_P \sqrt{\epsilon_V} \, $ from horizon exit to the end of inflation~\cite{Baumann:2014nda}. Satisfactory inflation requires an $N_* \sim O(50) - O(60)$ number of e-folds and this can be achieved in various models, albeit with some amount of fine-tuning of the potential. It is important to note that we can implement any inflationary ``Hilltop" model in our construction.

\subsection{Cosmological correlators}

It is also possible to compute cosmological correlators within our proposal, using the $AdS/CFT$ correspondence. Given a specific Euclidean geometry (i.e. a ``wineglass" wormhole), the first step of the computation corresponds to evaluating the corresponding Feynman-Witten diagrams~\cite{Witten:1998qj,Gubser:1998bc}, the correlator points placed on the $\tau=0$ slice and being connected with all possible bulk propagators and interaction vertices. One can equivalently use the wavefunction at $\tau = 0$ and determine the correlators via the schematic formula $\int D \phi \, |\Psi_{\tau = 0}|^2 \, \phi(0, \vec{x}_1) ... \phi(0, \vec{x}_n) $. The subsequent Lorentzian evolution of these correlators is governed by the precise form of the scalar potential on the right of the dashed line in Fig.~\ref{fig:PotentialHistory}. Due to the inherent time dependence of the cosmological background they are most appropriately computed using an in-in formalism~\cite{Weinberg:2005vy} (the Euclidean evolution determines the in-state). Another option is to evolve the $\tau = 0$ wavefunction into the Lorentzian future and use the evolved wavefunction $\Psi_{t}$ to compute the correlators at any later time.

\section{Wavefunctions and the probability measure}

\subsection{Issues with the no-boundary and Vilenkin proposals}

In the context of quantum cosmology, there have been several interesting proposals put forward on how to compute the wavefunction of the universe in the minisuperspace approximation (see eqn.~\eqref{minisuperansatze}), when the Wheeler-DeWitt wavefunction obeying the Hamiltonian constraint
\be\label{WDWequation}
\left[\frac{\partial^2}{\partial A^2} -  \frac{\partial^2}{\partial \tilde{\phi}^2} + \left( \frac{12 \pi^2}{\kappa } \right)^2 \left( e^{6 A} \tilde{V}(\tilde{\phi}) - e^{4 A} + \tilde{Q}^2 \right) \right] \Psi = 0 \, 
\ee
is well defined (we use the variable $A = \log a$ to avoid normal ordering ambiguities~\cite{doi:10.1142/1190} and define $\tilde{\phi} = \phi/M_{Pl}, \,  \tilde{V} = \kappa V/3  \, , \tilde{Q}^2 = \kappa Q^2/3$). One needs to supplement this equation with boundary conditions, the two main proposals being Hartle-Hawking's (No Boundary - NB)~\cite{Hartle:1983ai,Hartle:2007gi} and Vilenkin's (Tunneling - T)~\cite{Vilenkin:1982de,Vilenkin:1983xq} proposals. The NB proposal consists in extending the time evolution into the Euclidean regime in the far past and performing a sum over compact Euclidean geometries, thus avoiding the Lorentzian Big-Bang singularity. At the classical level this means that the Euclidean geometry smoothly caps off at some $\tau_{\text{initial}}$ (so that $a(\tau_{\text{initial}}) = 0 = \phi'(\tau_{\text{initial}})$), see Fig.~\ref{fig:wavefunctions} for an example.
In the tunneling proposal one demands a regular-bounded wavefunction consisting of outgoing modes at the boundary of superspace (this injects probability flux). Both proposals seem to suffer from some issues that are complementary in nature. In the semi-classical (WKB) approximation, the wavefunctions are found to behave in the oscillatory (Lorentzian) region $e^{2 A} \tilde{V}(\tilde{\phi}) \gg 1 \, , \tilde{Q}=0 \, $ as follows~\cite{doi:10.1142/1190,Lehners:2023yrj,Maldacena:2024uhs,Janssen:2020pii}
\bea\label{WKBwavefunctions}
\Psi_{NB}(A, \tilde{\phi}) \simeq {P_{NB}^{1/2}} \,  \Re \left( e^{ i S_L(A, \tilde{\phi})} \right) \, , \quad P_{NB} = e^{- S_E(\tilde{\phi})} \, ,  \nn \\
\Psi_{T}(A, \tilde{\phi}) \simeq {P_{T}^{1/2}} \left( e^{- i S_L(A, \tilde{\phi})} \right) \, , \quad {P_{T}} = e^{+ S_E(\tilde{\phi})} \, , \nn \\
S_E(\tilde{\phi}) = - \frac{ 8 \pi^2}{\kappa \tilde{V}(\tilde{\phi})}  \, , \quad  S_L(A, \tilde{\phi}) \simeq \frac{8 \pi^2 (e^{2 A} \tilde{V}(\tilde{\phi}) - 1)^{3/2}}{ \kappa \tilde{V}(\tilde{\phi})}  \,  . \nn \\
\eea
These formulae are valid assuming a slowly varying scalar potential i.e. when the slow roll approximation given by eqn.~\eqref{slowrollpar} holds. One observes that the oscillating part is related to the Lorentzian on-shell action $S_L$ and that the no-boundary wavefunction is real (stemming from the CPT symmetry of the Hartle-Hawking state). The prefactor $P^{1/2}$ is determined by a WKB matching procedure and depends exponentially on the Euclidean on-shell action $S_E(\tilde{\phi})$ of the De-Sitter instanton with an $S^4$ topology. The sign in the exponent in the two cases is opposite though, due to the different boundary conditions in the space of three geometries in the Euclidean past. The probability measure is then computed from $P = |\Psi|^2$ and one observes from eqns.~\eqref{WKBwavefunctions} that the Hartle-Hawking proposal exponentially peaks at the smallest (positive) value of the potential~\footnote{We should emphasize that this issue crucially rests upon the fact that the on-shell action of the Euclidean De-Sitter instanton is \emph{negative}. The value of the inflaton is also set by the sphere radius at horizon crossing.}. This leads to the smallest possible number of inflationary e-folds, in clash with observations. 
On the other hand the Vilenkin proposal fares well in this respect, allowing a large period of inflation due to the opposite sign in the exponent, making the wavefunction to peak at high values of the potential.
This sign though, is also crucial in determining the properties of fluctuations around these homogeneous and isotropic saddles. One then discovers an opposite troublesome behaviour ---that is the saddle points with the $(+)S_E$ sign are associated with enhanced fluctuations, and vice versa. This means that the Hartle-Hawking wavefunction defines the correct Euclidean or ``Bunch-Davies" vacuum and predicts the correct spectrum of primordial perturbations with a Gaussian suppression factor~\cite{doi:10.1142/1190,Lehners:2023yrj,Maldacena:2024uhs}. On the other hand Vilenkin's proposal is problematic in this regard, since fluctuations seem to grow and perturbation theory to break down~\footnote{It has been argued though, that imposing Robin boundary conditions for the modes at vanishing scale factor could potentially cure this problem~\cite{Vilenkin:2018dch}.}.

\subsection{Properties of our $AdS$ wormhole proposal}

In our proposal the WDW wavefunction also obeys eqn.~\eqref{WDWequation}. In the semi-classical WKB region $a^2 \tilde{V}(\tilde{\phi}) \gg  1$, it admits the same slow-roll approximation for the potential and hence the oscillatory part of the wavefunction behaves as in the no-boundary proposal (compare them in Fig.~\ref{fig:wavefunctions}). The difference concerns the real/Euclidean part of the wavefunction, now determined by the $EAdS$ boundary conditions and the resulting Euclidean on-shell action at the semi-classical level, that reads
\bea\label{onshellaction}
S^{\text{on-shell}}_E = 4 \pi^2 \int_{UV}^0 d \tau \left(\frac{2 Q^2}{a^3} - a^3 V  \right) + S_{GH}^{UV} + S_{c.t.}^{UV} \, , \nn \\
\eea
where the UV boundary ($\tau \rightarrow - \infty$) contribution contains the Gibbons-Hawking $S_{GH}^{UV}$ as well as boundary counterterms $S_{c.t.}^{UV}$ that one needs to add in order to perform holographic renormalization and render it finite for spaces containing an asymptotic $EAdS$ boundary (this is the ``UV-region" of the geometry). The particular four dimensional $EAdS$ example with an $S^3$ boundary was treated in detail in~\cite{Jafferis:2011zi,Taylor:2016kic,Ghosh:2018qtg}. The renormalised on-shell action for a solution at the minimum of the potential is positive $S^{\text{on-shell}}_{EAdS} = - {8 \pi^2 }/{\kappa \tilde{V}_{\text{min}}}$ (the unitary holographic dual CFT obeys the $F$-theorem). We then split the integral for the action of ``wineglass" $EAdS$ wormholes in two pieces, see Fig.~\ref{fig:winewormhole}. The first concerns the anti-friction region and is bounded from below by the positive action of $EAdS$ at $\tilde{V}_{\text{min}}$ ($EAdS$ is the most symmetric configuration within our ansatze~\eqref{minisuperansatze}). This piece to first approximation does not depend on $\tilde{V}_0$, but on the shape of the potential in the anti-friction region. The second piece concerns the friction region and can be evaluated analytically in the two complementary limits $a_{min} \ll a_{max}$ and $a_{min} \simeq a_{max}$~\footnote{See the supplemental material for details.}. The first limit tends to the no-boundary proposal result for $\tilde{Q} =0$ as expected. For small non zero $\tilde{Q}$, the action $S_E(\tilde{V_0})$ exhibits a local maximum as a function of $\tilde{V}_0$ and the probability $P(\tilde{V}_0) = |\Psi|^2 \simeq  e^{- S_E}$ maximises towards either increasing or decreasing $\tilde{V}_0$ as long as it obeys the bounds set by the potential $\tilde{V}_{ms} \leq \tilde{V}_0 \leq \tilde{V}_{max} $ ($\tilde{V}_{ms}$ is the local positive minimum at the end of inflation and $\tilde{V}_{max}$ is the positive maximum of the potential). In the second complementary limit we again find a similar behaviour. Depending on the precise details of the potential in our model, the largest possible inflationary period can be favored.

\section{Discussion}

In this work we proposed a new type of wavefunction of the universe and its semiclassical limit corresponding to a ``wineglass" shaped $EAdS$ half-wormhole, that sets appropriate initial conditions for inflation. Our proposal can evade the issues afflicting other well known proposals, leading to a well behaved probability favoring a long lasting inflationary period due to the $EAdS$ boundary conditions in the far past, and a reasonable spectrum of fluctuations due to its similarity with the no-boundary proposal in the transition region into Lorentzian signature, see Fig.~\ref{fig:wavefunctions}.

It would be interesting to perform a thorough analysis of cosmological perturbations in our setup and find whether or not it leads to any deviations from the usual inflationary paradigm. An analysis of the WDW equation using Picard-Lefschetz theory could also elucidate further the properties of our proposal, see~\cite{DiazDorronsoro:2017hti,Feldbrugge:2018gin,DiazDorronsoro:2018wro} for some works in this direction. It is also straightforward to add additional matter and radiation densities to the Friedmann equations~\eqref{EEOMs}, that are important to describe the evolution of a more realistic model of the universe. We would also like to estimate the probability of the metastable $dS$ vacuum to decay back to $AdS$ ---we expect this to be extremely small for a slow roll potential, such as the one drawn in Fig.~\ref{fig:PotentialHistory} (coupled to the fact that gravitation has the tendency to stabilise false vacua~\cite{Coleman:1980aw}). Finally, it is natural to ponder whether potentials with characteristics as depicted in Fig.~\ref{fig:PotentialHistory} can arise in a top-down string theory construction.

\begin{acknowledgments}

We wish to thank Elias Kiritsis, Juan Maldacena, Mark van Raamsdonk and the rest of the physics group at UBC for discussions and comments. We also acknowledge useful discussions during the \href{https://www.kitp.ucsb.edu/activities/hartle-c24}{Jim Hartle’s Legacy conference} and the hospitality of KITP, where part of this work was undertaken.

\noindent The research of P.B. is supported in part by the Natural Sciences and Engineering Research Council of Canada. P.B acknowledges support by the Simons foundation. This research was also supported in part by the grant NSF PHY-2309135 to the Kavli Institute for Theoretical Physics (KITP).

\end{acknowledgments}

\bibliography{worminflation}

\onecolumngrid
\appendix

\section*{Supplemental Material \\ Properties of the solutions and the on-shell action}

In this supplemental material, we provide some further details on the general properties of the solutions and the evaluation of the on-shell action for our model of ``wineglass" $EAdS$ (half)-wormholes described in the main text.

In order not to clutter the formulae, we rescale to Planck units $\tilde{\phi} \equiv \phi/M_{Pl}, \,  \tilde{V} \equiv \kappa V/3, \,  \tilde{Q}^2 \equiv \kappa Q^2/3 = \kappa q^2/6 f_\alpha^2$. According to Fig.~\ref{fig:PotentialHistory}, we also assume a scalar potential with a negative maximum $\tilde{V}_{\tau = - \infty} = \tilde{V}(\tilde{\phi} = 0)$, a global negative minimum $\tilde{V}_{min}$, a positive maximum $\tilde{V}_{max}$ and a positive metastable minimum $\tilde{V}_{ms}$. These are all parameters that are fixed in a specific model. The Euclidean motion in $\tau \in (-\infty, 0)$ is performed in the inverted potential $-\tilde{V}(\tilde{\phi})$. The potential at $\tau = 0 $ is $\tilde{V}_0$, which is bounded by $\tilde{V}_{max} > \tilde{V}_0 \geq \tilde{V}_{ms} > 0$.

We first integrate the scalar inflaton equation in~\eqref{EEOMs}
\be\label{inflatonintegrated}
\frac{1}{6} \tilde{\phi}'^2 -   \tilde{V}(\tilde{\phi})  -  W_{\text{friction}}(\tau) = -  \tilde{V}_{\tau = - \infty} =  - W_{\text{friction}}^{\text{total}}(0) -  \tilde{V}_0 \, ,  \qquad W_{\text{friction}}(\tau) = -   \int_{-\infty}^\tau d \tilde{\tau} \frac{a' \tilde{\phi}'^2}{a} \, ,
\ee
where $W_{\text{friction}}(\tau)$ is the work done by the friction up to time $\tau$ for the motion in the effective potential $- \tilde{V}(\tilde{\phi})$. We then split the motion and the Euclidean time integral in eqn.~\eqref{onshellaction} in two regions as seen in Fig.~\ref{fig:winewormhole}. The first is for $\tau \in (-\infty, \tau_{\text{min}})$, where $a(\tau_{\text{min}}) = a_{\text{min}}$ (region of anti-friction) and the second is the region $\tau \in (\tau_{\text{min}}, 0)$ (region of friction).
The total work of the friction also splits into a positive and a negative contribution $W_{\text{friction}}^{\text{total}} = W_{\text{anti-friction}}^{+} + W_{\text{friction}}^{-}$, in the two corresponding regions.

Next we analyze the second (constraint) equation of~\eqref{EEOMs}, at $\tau=0$, when $a'= \tilde{\phi}' = 0$. We find the cubic equation
\be\label{cubic}
\tilde{V}_0 x^3 - x^2 + \tilde{Q}^2 = 0 \, , \qquad \Delta = \tilde{Q}^2 (4-27 \tilde{Q}^2  \tilde{V}^2_0 ) \, , \qquad x = a_0^2 \, .
\ee
For $\Delta > 0$ (this is the phenomenologically viable case of relatively small axion charge), this equation has three roots, since $\tilde{V}_0>0$ two of them are positive and one negative. They are explicitly given by
\be\label{cubicsln}
\cos \theta = 1 - \frac{27}{2} \tilde{Q}^2  \tilde{V}^2_0  \, , \quad \theta \in (0,\pi] \, , \qquad x_n = \frac{1}{3 \tilde{V}_0} \left(1 + 2 \cos \left(\frac{\theta - 2 \pi n}{3} \right) \right) \, , \quad n = 0 , 1, 2 \, .
\ee
Using these solutions in the first equation of~\eqref{EEOMs}, we find that only the largest root ($n=0$) satisfies the condition $a''(0) < 0$ so that $a_0 = a_{max}$. We also observe that $a_{max}$ satisfies the condition in eqn.~\eqref{boundamax}, so that it is directly related to the inverse of $\tilde{V}_0$.

The third step involves repeating this analysis for $a({\tau_{\min}}) = a_{min}, \, a'({\tau_{\min}}) = 0$. In this case, $\phi'({\tau_{\min}}) \neq 0$, but we can borrow the results in eqns.~\eqref{cubic} and~\eqref{cubicsln} by simply substituting $\tilde{V}_0 \rightarrow \tilde{V}_{\tau_{min}} - \phi'^2_{\tau_{min}}/6 = \tilde{V}_0 + W_{\text{friction}}^{-}$. Now depending on the details of the shape of the potential and where the anti-friction transitions into friction, this could either be a positive or negative number. Since we wish to analyse the cases where the friction regime starts early for small $\tilde{\phi}$ (that is for $\tilde{V}_{\tau_{min}}<0$), we have to pick the $x_2$ solution in \eqref{cubicsln} that is now positive. This then determines $x_2 = a_{min}^2$
as a function of the parameters of the model. In fact the earliest that the friction regime can start, is such that the particle's kinetic energy when it reaches $\tilde{V}_{min}$ is almost zero.

We finally discuss the evaluation of the on-shell action. The part of the integral~\eqref{onshellaction} in the anti-friction region is manifestly positive (and bounded from below by the action of $EAdS$ at $\tilde{V}_{\text{min}}$) ---in the context of holography it describes a renormalization group flow driven by a relevant deformation that is skipping the $EAdS$ fixed point at the minimum of the potential. The second piece concerns the friction region and contains both a positive and a negative contribution depending on the sign of the potential $\tilde{V}(\tilde{\phi})$. This can be evaluated analytically in the following two complementary limiting cases: 

\begin{itemize}
   
\item The first limit describes wormholes with $a_{\text{min}} \ll a_{\text{max}}$, so that the contracting $EAdS$ region is attached with a thin neck to the expanding (almost Euclidean De-Sitter) region, see Fig.~\ref{fig:wavefunctions}. This approximation is valid in the case that $|W_{\text{friction}}^{-}| \gg  \tilde{V}_0$. In this limit, near $\tau_{min}$, there is a very narrow region where $a' \simeq 0$ (where the friction term can be neglected to first order). This is a ``thin wall" (small $\Delta \tau$) transition region, similar to the one studied in~\cite{Coleman:1980aw}, where the scalar field changes abruptly its value from $\tilde{\phi}_{\tau_\text{min}}$ to a value close to $\tilde{\phi}_0$ and where the scale factor is approximately a constant very close to $a_{\text{min}} $. This transitions abruptly to an outer ``thick" (large $\Delta \tau$) region where $a'$ cannot be neglected, but for which the potential becomes approximately constant (a``slow roll" region). In this ``thick" region, the evolution of the scale factor approximates that of Euclidean De-Sitter. The end of this region is when the scale factor reaches $a_{\text{max}}$. In the ``thin-wall" region, the on-shell action has a contribution
\be\label{thinaction}
S_E^{\text{thin-wall}} \simeq  \frac{12 \pi^2}{\kappa} \int_{thin} \, d \tau \, \left( \frac{2 \tilde{Q}^2}{a_{\text{min}}^3} - a_{\text{min}}^3 \tilde{V}(\tilde{\phi}) \right)  \simeq \frac{2 4 \pi^2 \tilde{Q}^2 }{\kappa {a_{\text{min}}^3}} \Delta \tau_{thin}    \, ,
\ee
that is manifestly positive and independent of $\tilde{V}_0$ to first approximation.
One can also evaluate the on-shell action~\eqref{onshellaction} in the outer ``thick" region, where $a' \neq 0$, by solving the second equation in~\eqref{EEOMs} with an approximately constant $\phi \simeq \tilde{\phi}_0, \, \tilde{V} \simeq \tilde{V}_0 \sim 1/a_{max}^2$ and small $\tilde{Q}$, to find 
\begin{align}\label{outeraction}
S_E^{\text{outer-thick}} \simeq  \frac{12 \pi^2 \tilde{Q}^2}{\kappa} \left( \frac{\sqrt{ 1- a^2_{min} \tilde{V}_0 }}{a_{min}^2} + \tilde{V}_0 \tanh^{-1} (\sqrt{ 1- a^2_{min} \tilde{V}_0 })   \right)   - \frac{4 \pi^2}{\kappa \tilde{V}_0} \left( 1- a^2_{min} \tilde{V}_0 \right)^{3/2} \, = \, \nn \\
= \frac{12 \pi^2 \tilde{Q}^2}{\kappa} \left(\frac{1}{a^2_{min}} - \tilde{V}_0 \log \frac{a_{min} \sqrt{\tilde{V}_0} }{2}  \right) -   \frac{4 \pi^2}{\kappa \tilde{V}_0} \, + \, O(a_{min}^2 \tilde{V}_0) \, . 
\end{align}
As a check we observe that for $\tilde{Q} = 0$ and as $a_{min}/a_{max} \rightarrow 0$, the action tends to the one given in the no-boundary proposal~\cite{Hartle:1983ai,doi:10.1142/1190} (half the De-Sitter instanton) as we intuitively expect, since then the Euclidean space would correspond to a smooth half $S^4$ and the $EAdS$ asymptotic region would completely detach. Considering then the action as a function of $\tilde{V}_0$, it gets attracted to the smallest possible value of $\tilde{V}_0$ (the positive metastable minimum $\tilde{V}_{ms}$) ---this is the known issue of the no-boundary proposal. On the other hand, if we keep $\tilde{Q}, a_{min}$ small but non-zero and consider again the action as a function of $\tilde{V}_0$, we find that there is an unstable maximum of the action for $\tilde{V}_0 = \tilde{V}_*$. This means that the probability $P = |\Psi|^2 = e^{- S_E(\tilde{V}_0)}$ is attracted and maximises either for the smallest or largest possible values of $\tilde{V}_0$, as long as the bound $\tilde{V}_{ms} \leq \tilde{V}_0 \leq \tilde{V}_{max}$ is satisfied. Of course if the unstable maximum of the action $\tilde{V}_*$ is smaller than $\tilde{V}_{ms}$, the only possibility left is a runaway behaviour towards the largest possible value $\tilde{V}_{max}$ ---the wanted feature giving rise to the biggest possible number of inflationary e-folds. On the other hand, since at large values of $\tilde{V}_0$ our assumption $a_{min} \ll a_{max}$ or $|W_{\text{friction}}^{-}| \gg  \tilde{V}_0$ could potentially break down, we also need to understand the opposite limit, where $a_{min}$ is comparable to $a_{max}$.

\item We now study the opposite case where $a_{min}$ is comparable to $a_{max}$. We can treat analytically the limiting case when $a'/a \simeq 0$ in an inner ``thick" region, while the previous outer region where $a' \neq 0$ has now shrunk and is very ``thin". We now find that most of the action integral of eqn.\eqref{onshellaction} is concentrated in the inner ``thick" region
\bea\label{innerthickaction}
S_E^{\text{inner-thick}} \simeq \frac{12 \pi^2}{\kappa} \int_{thick} \, d \tau \, \left( \frac{2 \tilde{Q}^2}{\overline{a}^3} -\overline{a}^3 \tilde{V}(\tilde{\phi}) \right) \, \simeq \nn \\
\simeq   \frac{1 2 \pi^2 }{\kappa} \int_{\tilde{\phi}_{\tau_{min}}}^{\tilde{\phi_0}} d \tilde{\phi} \left( \frac{2 \tilde{Q}^2 / \overline{a}^3  - \overline{a}^3   \tilde{V}(\tilde{\phi})}{\sqrt{6 ( \tilde{V}(\tilde{\phi}) - C)}} \right) \, ,
\eea
where we used eqn.~\eqref{inflatonintegrated}, in the case of small $a'/a$ ($\overline{a}$ can be taken to be the average of $a_{min}$ and $a_{max}$). Moreover $C$ is a constant for which $V_{min} < C < V_{\tau_{min}} $ (since the motion under friction starts with some non zero kinetic energy picked up during anti-friction).
On the other hand the integral in the outer ``thin" region, where the scalar potential is approximately constant is given again by the first line of eqn.~\eqref{outeraction}, but vanishes in the limit $a_{min} \simeq a_{max}$ and we can therefore neglect it to first approximation.

The integral in eqn.~\ref{innerthickaction} depends on $\tilde{V}_0$ through its end-point at $\tilde{\phi}_0$. Expanding the potential near that point as $\tilde{V}(\tilde{\phi}) = \tilde{V}_0(1 - \epsilon_{\tilde{V}} \tilde{\phi})$, with $\epsilon_{\tilde{V}} \ll 1$ a positive slow-roll parameter we find
\be
S_E^{\text{inner-thick}}(\tilde{V}_0) \simeq  \sqrt{\frac{2}{3}}\frac{4 \pi^2 \sqrt{\tilde{V}_0-C}   }{\kappa \overline{a}^3 \epsilon_{\tilde{V}} \tilde{V}_0} \left(- 6 \tilde{Q}^2 + \overline{a}^6 (2 C + \tilde{V}_0) \right) \, .
\ee
We then replace $\overline{a} = r/\sqrt{\tilde{V}_0}$ with $r$ and $O(1)$ number and minimise $S_E^{\text{inner-thick}}(\tilde{V}_0)$.
As for the opposite case $a_{min} \ll a_{max}$, we find again an unstable maximum. Once more if this maximum of the action is below $\tilde{V}_{ms}$, the only possibility is the runaway towards the largest allowed number of $\tilde{V}_0$ that is $\tilde{V}_{max}$, giving rise to the maximum possible inflationary period.

\end{itemize}

\end{document}